\documentclass[aps,pra,twocolumn,preprintnumbers,amsmath,amssymb,showpacs,nofootinbib,secnumarabic]{revtex4}

\usepackage{graphicx}
\usepackage{dcolumn}
\usepackage{bm}
\usepackage{epsfig,hyperref}
\usepackage{color}
\usepackage{relsize}
\usepackage{accents}
\usepackage{enumerate}

\usepackage[caption=false]{subfig}
\def\sct{0.45} 

\def\art{paper}

\def\jrn#1#2#3#4#5#6{#3 \textbf{#4}, #5 (#6)} \def\andd{and } 

\def\scn#1#2{\section{#1}\lb{#2}}
\def\sscn#1#2{\subsection{#1}\lb{#2}}
\def\bfl{\begin{flushleft}}
\def\efl{\end{flushleft}}
\def\bfr{\begin{flushright}}
\def\efr{\end{flushright}}
\def\bc{\begin{center}}
\def\ec{\end{center}}
\def\bse{\begin{subequations}}
\def\ese{\end{subequations}}
\def\baa#1{\begin{array}{#1}}
\def\eaa{\end{array}}
\def\bw{\begin{widetext}}
\def\ew{\end{widetext}}
\def\nn{\nonumber }
\def\lb#1{\label{#1}}
\def\bit{\begin{itemize}}
\def\eit{\end{itemize}}
\def\bco{}
\def\bcs{\begin{cases}}
\def\ecs{\end{cases}}

\def\densnnorm{\hat{\rho}}
\def\densnorm{\hat{\rho}^\prime}

\def\nii{\ell}
\def\nhc{\gamma}
\def\poiss{\Pi} 

\def\av#1{\langle #1 \rangle}
\def\avo#1{\av{#1}'}

\def\wigdens{W_\rho}

\newcommand{\be}{\begin{equation}}
\newcommand{\ee}{\end{equation}}
\newcommand{\ba}{\begin{eqnarray}}
\newcommand{\ea}{\end{eqnarray}}

\newcommand{\lan}{\langle}
\newcommand{\ran}{\rangle}

\newcommand{\p}{\partial}

\makeatletter
\DeclareRobustCommand{\cev}[1]{%
  \mathpalette\do@cev{#1}%
}
\newcommand{\do@cev}[2]{%
  \fix@cev{#1}{+}%
  \reflectbox{$\m@th#1\vec{\reflectbox{$\fix@cev{#1}{-}\m@th#1#2\fix@cev{#1}{+}$}}$}%
  \fix@cev{#1}{-}%
}
\newcommand{\fix@cev}[2]{%
  \ifx#1\displaystyle
    \mkern#23mu
  \else
    \ifx#1\textstyle
      \mkern#23mu
    \else
      \ifx#1\scriptstyle
        \mkern#22mu
      \else
        \mkern#22mu
      \fi
    \fi
  \fi
}

\makeatother

\begin{document}

\preprint{\small Int. J. Mod. Phys. B \textbf{32}, 1850276 (2018)   
\quad 
[\href{https://doi.org/10.1142/S0217979218502764}{DOI: 10.1142/S0217979218502764}]
}

\title{Phase space formulation of density operator for non-Hermitian Hamiltonians
and its application in quantum theory of decay
}

\author{Ludmila Praxmeyer}
\email{lpraxm@gmail.com}
\affiliation{Institute of Systems Science, Durban University of Technology,
P. O. Box 1334, Durban 4000, South Africa}

\author{Konstantin G. Zloshchastiev}
\email{https://bit.do/kgz}
\affiliation{Institute of Systems Science, Durban University of Technology,
P. O. Box 1334, Durban 4000, South Africa}

\begin{abstract}
The Wigner-Weyl transform and phase space formulation
of a density matrix approach are 
applied to a non-Hermitian model which is quadratic in positions and momenta.
We show that in the presence of a quantum environment or reservoir,
mean lifetime and decay constants of quantum
systems do not necessarily take arbitrary values, but could become functions of energy eigenvalues
and have a discrete spectrum.
It is demonstrated also that a constraint upon mean lifetime and energy appears,
which is used to derive the resonance conditions at which long-lived states occur.
The latter indicate that quantum dissipative effects do not always
lead to decay but, under certain conditions, can support stability of a system.
\end{abstract}

\maketitle

\scn{Introduction}{s:intr}

When studying open quantum systems \cite{fbook,bpbook},
it is usually implied that decay constants and mean lifetimes of states
take their numerical values from a continuous spectrum;
at least, no systematic theory for describing forbidden regions or discrete spectra of those
values has been yet proposed, to the best of our knowledge.
Analytical approaches and numerical packages use this continuity, often implicitly,
because it allows to drastically simplify calculations \cite{fgr78,cl83}.
Nevertheless, there has been no formal proof of the non-existence of those phenomena to date either.
On the other hand, experimental data exist whose explanation
could invoke discrete spectra of mean lifetimes,
such as experiments with quantum dots and nanocrystals \cite{wmm13,wmm15,ml15}.

Another long-standing problem
is the existence of long-lived states
in open quantum systems.
In conventional theoretical models,
wavefunctions or density matrices
of systems with sinks or sources usually experience some kind
of decay or unbound increase behavior, usually at an exponential rate with time.
On the other hand, realistic quantum systems are mostly open ones, yet many of
them clearly have long-lived or even
stationary states, one striking example being photobiological and light-harvesting complexes,
where laser-pulse femtosecond photon echo spectroscopy experiments
reveal the long-lived exciton-electron quantum coherence in photosynthetic reaction centers, even
at room temperatures
\cite{sbs97,ecr07,lcf07,ran14}.
This poses the question of
how to describe those
open quantum systems whose states can be not only decaying or unstable,
but also
long-lived, depending on conditions.

In this \art, we demonstrate a universal mechanism through which:
(i) decay constants and mean lifetimes
can become quantized so that their values acquire discrete spectra,
(ii) open quantum systems do not have all states with decaying or unstable behavior,
but can also have some states which are long-lived.

The contents of this paper are as follows.
In section~\ref{s:bm},
we formulate the non-Hermitian (NH) quantum harmonic oscillator model,
which can describe a wide range of dissipative phenomena, exactly
or in the harmonic approximation.
In section~\ref{s:ee},
we provide a brief account of the density matrix approach for the
systems with non-Hermitian Hamiltonians.
In section~\ref{s:ps3},
the Wigner-Weyl transform and phase space formulation
of a density-matrix approach for non-Hermitian systems are presented
and adapted for our case.
In section~\ref{s:cases},
we illustrate the phase-space density matrix approach by solving analytically
the evolution equation for some special cases, and consider the physical properties thereof.
Discussions and conclusions are given in section~\ref{s-con}.

\scn{The model}{s:bm}
%
We work within the frameworks of the non-Hermitian Hamiltonian approach
to quantum dissipative systems,
which have become popular since the classical works devoted to a theory of
many-particle systems
\cite{suu54,fesh1}.
According to this method, one assumes that
the anti-Hermitian part arises in a Hamiltonian as a
result of the interaction of an otherwise conservative system with its
environment or reservoir \cite{fbook,bpbook,nimrod}.
Despite the long history of the field,
the core formalism of non-Hermitian quantum dynamics is still a subject of active research from the viewpoint of 
the density operator approach
\cite{sz13,sz14,sz14cor,kar14,sz15,z15,sg16,zfx17} and in connection with nonlinear quantum dynamics \cite{zrz17}.


Let us consider an open quantum system
where the effect of its environment is described by introducing
an anti-Hermitian part of its Hamiltonian.
The resulting non-Hermitian Hamiltonian operator can be written in the
form
\be
\hat{\cal H}=
\hat H -i \hat\Gamma \;,
\label{e:nhhtot}
\ee
where both $\hat H$ and $\hat\Gamma$ are Hermitian operators;
$\hat\Gamma$ is often called the \textit{decay rate operator}, in order to emphasize its relation to dissipative environment \cite{sz13}.
If anharmonic effects are neglected
then we can adopt the harmonic approximation for both operators:
\ba
\hat H
=
\frac{\hat{p}^2}{2m} +\frac{m\omega^2}{2}
\hat{q}^2
,
\ \
\hat\Gamma
=
\frac{\alpha}{2} \hat{p}^2
+\frac{\beta}{2} \hat{q}^2
+ \frac{\gamma }{2}
,
\lb{mbhamg}
\ea
where we denoted the operators $\hat{q} = (\hat{a}^\dagger +
\hat{a})\sqrt{\hbar/(2 m \omega)}$ and $\hat{p} = i \sqrt{m
\omega\hbar/2}(\hat{a}^\dagger - \hat{a})$, $\hat{a}^\dagger$ and
$\hat{a}$ being the creation and annihilation operators,
respectively.
Alternatively,
one can write
\ba
\hat H
&=&
\hbar \omega
\left(
\hat{a}^\dagger \hat{a}
+ \frac{1}{2}
\right)
,
\nn \\
\hat\Gamma
&=&
\frac{\nhc_+}{2}
\left(
\hat{a}^\dagger \hat{a}
+ \frac{1}{2}
\right)
-
\frac{\nhc_-}{4}
\left(
\hat{a}^{\dagger 2} + \hat{a}^2
\right)
+
\frac{\gamma}{2}
,
\ea
where $\nhc_\pm = \hbar \omega ( \alpha m \pm \beta/(m \omega^2))$,
which is a more convenient form for condensed matter models.
The parameters $m$, $\omega$,
$\alpha$ and $\beta$ are positive, and $\gamma$ is real-valued; the
latter three  will hereafter be referred as the non-Hermitian parameters.

Systems with Hamiltonians 
quadratic in position and momenta are ubiquitous in physics,
because the harmonic approximation can be applied to nearly any system
which allows a Hamiltonian description,
therefore, we deliberately do not specify a model here.
In models of type (\ref{mbhamg}), non-Hermitian parameters
can naturally emerge as a
result of a extension of conventional (Hermitian)
harmonic oscillator's parameters into a complex domain -- to take
into account various dissipative phenomena, which occur as a result
of interaction of an oscillator with its environment. If a
non-Hermitian model is used as an effective description of quantum
many-body systems, then the non-Hermitian parameters are usually related to
dissipative or radiative corrections to scattering and tunneling
amplitudes of an underlying ``exact'' system. 
Otherwise, their
physical meaning is determined from their context, some examples to
be found in Refs.
\cite{sz14,kar14,z15,z16,z17adp,li17} and Ch. 6 of \cite{bpbook}.

Performing scale transformations
with respect to the characteristic values $q_c =
\hbar/p_c =
\sqrt{\hbar/(m
\omega)}$ and $E_\omega =
\hbar
\omega$,
and introducing
notations
\be
\tilde\alpha = \alpha m ,\ \tilde\beta = \beta/(m \omega^2),\  \tilde\gamma = \gamma/E_\omega
,
\ee
we can obtain a dimensionless expression of the Hamiltonian (\ref{mbhamg}).
It reads
\ba
\hat H
&=&
\frac{\hat p^2}{2}
+
\frac{\hat q^2}{2}
,\ \
\hat\Gamma
=
\frac{\alpha}{2}  \hat p^2
+
\frac{\beta}{2}  \hat q^2
+
\frac{\gamma}{2}
,
\lb{mbhamg-qp}
\ea
where
we have omitted tildes for a sake of brevity:
from now on we assume that values with dimensionality of
distance, momentum and energy are
measured in units $q_c$, $p_c$ and $E_\omega$, respectively,
while the non-Hermitian parameters $\alpha$, $\beta$ and $\gamma$
are measured in units $m^{-1}$, $m \omega^2$ and $E_\omega$, respectively.
Alternatively, we can also write 
\ba
\hat H
&=&
\hat{a}^\dagger \hat{a}
+ \frac{1}{2}
,
\nn \\
\hat\Gamma
&=&
\frac{\nhc_+}{2}
\left(
\hat{a}^\dagger \hat{a}
+ \frac{1}{2}
\right)
-
\frac{\nhc_-}{4}
\left(
\hat{a}^{\dagger 2} + \hat{a}^2
\right)
+
\frac{\gamma}{2}
,
\ea
where $\nhc_\pm = \alpha\pm\beta$.

\scn{Master equation for non-Hermitian systems}{s:ee}

In order to study the dynamics of the system defined in the previous
section, one must develop a formalism which deals with non-Hermitian Hamiltonians.
We will be interested in the most general types of quantum
dynamics that involve not only pure but also mixed states,
therefore we will follow the reduced density matrix approach widely popular in
a theory of open quantum systems \cite{fbook,bpbook}.

Upon introducing a non-normalized 
density matrix of a subsystem,
\be
\densnnorm
=\sum_k{\cal P}_k|\Psi^k \rangle\langle\Psi^k |
,
\ee
where ${\cal P}_k$ are the probabilities of the states $(|\Psi^k \rangle,\langle\Psi^k |)$,
the quantum dynamics of non-Hermitian systems can be described in terms of the evolution equation,
\begin{equation}
\partial_t {\densnnorm} =-\frac{i}{\hbar}\left[\hat H, \densnnorm\right]-
\frac{1}{\hbar}
\left\{\hat{\Gamma},\densnnorm \right\}
\;,
\label{e1}
\end{equation}
where $[ , ]$ and $\left\{ , \right\}$ denote the commutator and
anticommutator, respectively \cite{fbook}.
In the context of the theory
of open quantum systems,
$\densnnorm$ is the reduced density operator, which  effectively describes the original
subsystem (with Hermitian Hamiltonian) together with the effect of
environment (represented by anti-Hermitian and Hermitian corrections).

Upon taking a trace
of both sides of Eq.~(\ref{e1}), one obtains an evolution equation
for the trace of $\densnnorm $,
$
\partial_t{\rm Tr}\,\densnnorm
=-\frac{2}{\hbar}{\rm Tr}(\hat{\Gamma} \,\densnnorm )
$, 
which shows that the trace is not conserved in general,
therefore non-Hermitian dynamics does not preserve
the probability measure.
As was suggested in Ref.~\cite{sz14}, one is then led to the introduction
of a normalized density operator, defined as
\begin{equation}
\densnorm
 =\densnnorm /{\rm Tr}\,\densnnorm
,
\label{e:rho}
\end{equation}
that can be used in the calculation of quantum statistical averages
of operators,
$
\avo{\hat A} = \text{Tr} (\hat A \densnorm)
= \text{Tr} (\hat A \densnnorm)/\text{Tr} \densnnorm
=\av{\hat A}/\text{Tr} \densnnorm
$, 
where $\hat A$ being an operator related to a physical observable.
As a result of
Eqs. (\ref{e1}) and (\ref{e:rho}),
the normalized density
operator obeys
the following evolution equation:
\be
\partial_t {\densnorm} =
-\frac{i}{\hbar}\left[\hat H, \densnorm\right]
-\frac{1}{\hbar}\left\{\hat{\Gamma},\densnorm \right\}
+\frac{2}{\hbar}\densnorm \, {\rm Tr} (\hat{\Gamma}\densnorm)
,
\label{eq:dotrho}\end{equation}
which is both nonlinear and nonlocal, due to the last term.
The operator $\densnorm$
is bounded and
allows one to maintain a probabilistic interpretation
of the statistical averages of operators under non-Hermitian dynamics.
Nevertheless, the gain or loss of probability associated with
the coupling to sinks or sources are properly described
by the non-normalized density operator $\densnnorm$.
Thus,
one can use both $\densnnorm$ and $\densnorm$
(which would determine, respectively, non-sustainable and sustainable types of evolution),
depending on a physical context and boundary conditions, see Refs. \cite{z16,z17adp} for some examples.

\scn{Phase space formulation}{s:ps3}

Practical computation of a density matrix from Eq. (\ref{e1})
poses a
significant technical problem, especially when the Hilbert
space of a system is infinite-dimensional. One of the approaches that allows
the reduction of this problem to
a differential equation
is the Weyl-Wigner transform and phase space formulation of quantum
mechanics
\cite{Wigner,schleich,book-phasespace,tarasow}.
Thus, the phase-space approach must be extended for the non-Hermitian case,
which will be the next task that we address
here (some previous work in this direction was done in Refs. \cite{gr46,cv07,gs11,bc13,bc14,bcl15,ser15w}).

Following notations of
Ref. \cite{tarasow},
the evolution equation (\ref{e1}) can be written as
\ba
\frac{\partial}{\partial t} \hat\rho=\hat{H}\cdot\hat\rho-\frac{2}{\hbar}\hat{\Gamma}\circ\hat\rho
,
\label{em}
\ea
where 
Lee and
Jordan multiplications are given by
$
\hat{A}\cdot \hat{B}= \frac{1}{i\hbar}\big(\hat{A} \hat{B}-\hat{B} \hat{A}\big)
$
and
$\hat{A}\circ
\hat{B}=\frac{1}{2}\big(\hat{A} \hat{B}+ \hat{B} \hat{A}\big)
$, respectively.
Furthermore, we introduce a Weyl symbol of a non-normalized density operator:
$
\wigdens(p,q,t)=\frac{1}{2\pi\hbar}\int d\xi \lan
q-\xi/2 |\hat\rho(t)|q+ \xi/2\ran e^{\frac{i \xi
p}{\hbar}}
$,
which will be called the non-normalized Wigner density function.

It can be
checked that if $\hat{\Gamma} \not=0$ then, in general,
normalization condition,
$
\int \wigdens(p,q,t) dq dp = 1
$,
does not
hold.
Therefore, differences between the non-normalized
$\wigdens$
and normalized
$
\wigdens' = \wigdens/ \int \wigdens dq dp
$
Wigner density functions
are non-trivial indeed.
An expectation value of an operator $\hat{A}$ in a state
$\hat\rho$ equals to the phase-space average of a product of the
Weyl transform  of that operator, $A_W(p,q)$, and the Wigner
function of $\hat\rho$,
$
\av{\hat{A}} = \int \wigdens(p,q,t) A_W(p,q) dq dp = \avo{\hat{A}} \int \wigdens(p,q,t) dq dp
$.

Using the fact that  $\hat{A}$ and $\hat{B}$ are Weyl ordered
operators, we obtain the following relations between their
Weyl transforms:
\ba &&
{(\hat{A}\cdot \hat{B})_W}(p,q)=-\frac{2}{\hbar}A_W(p,q) \sin  \big( {  
\frac{1}{2} \hbar\poiss }
\big)B_W(p,q),~~~~~ \
\\ &&
{(\hat{A}\circ \hat{B})_W}(p,q)=A_W(p,q)\cos \big({  
\frac{1}{2} \hbar\poiss  }
\big)B_W(p,q)
,
\ea
where
$
\poiss=\cev\partial_{p}\vec{\partial}_q-\cev\partial_{q}\vec\partial_p$;
the operation $\exp{\left(-\tfrac{i}{2} \poiss\right)}$ is usually called
the Groenewold star product \cite{gr46}.
Substituting these relations into Eq. (\ref{em}), we find that Eq. (\ref{e1}) in a phase space formulation 
is given by
\ba
\frac{\partial}{\partial t} \wigdens
&=&
-
\frac{2}{\hbar}H_W(p,q) \sin \big({
\tfrac{1}{2}\hbar\poiss  }
\big)\wigdens
\nn\\&&
-
\frac{2}{\hbar}\Gamma_W(p,q)\cos \big({
\tfrac{1}{2} \hbar\poiss}
\big)\wigdens
,
 \label{er}
\ea
which is similar to the equations used in the previous works \cite{cv07,gs11,bc13,bc14,bcl15}.
For the Hamiltonian (\ref{mbhamg-qp}), this equation takes a
form
\ba
\frac{\partial}{\partial t} W
&=&
-(
\alpha
 p^2
+
\beta
 q^2
+ \gamma)
W
-
\left(
p \frac{\partial}{\partial q}
-
q \frac{\partial}{\partial p}
\right)
W
\nn\\&&
+
\frac{1}{4}
\left(
\alpha \frac{\partial^2}{\partial q^2}
+
\beta \frac{\partial^2}{\partial p^2}
\right)
W
,
\lb{e:evgen}
\ea
where we drop a subscript of $\wigdens$;
we also work in $\hbar = 1$ units from now on.
One can see
that
Eq. (\ref{e:evgen}) is a second order differential equation,
which can be treated by analogy with wave equations known from quantum mechanics.
The first term on the right-hand side of Eq. (\ref{e:evgen}) corresponds to a 
plain exponential decay (or growth, depending on signs of the coefficients);
its strength depends on a distance from the phase-space's origin $q = p = 0$.
The effective decay constant thus becomes a function of position and momentum:
$W \sim \exp{(- \gamma_\text{eff}\, t)}$,
$\gamma_\text{eff} = \gamma +\alpha  p^2 + \beta  q^2 $.
The second term comes from the Hermitian part of the Hamiltonian,
which is a quantum harmonic oscillator in this case; it leads to the rotation of
solutions with time, as expected. 
The third term is yet another non-Hermitian effect:
it describes transfer of probability in the phase space,
which is somewhat similar to heat transfer (if $\alpha$ and $\beta$ are both positive,
as in partial differential equations of an elliptic type),
otherwise it is a term with no simple classical analogue.

\scn{Special cases}{s:cases}

In this section we consider two special cases, which are related to
different symmetries in phase space, 
and whose spectra of interest take particularly simple forms,
which allows us to directly perform an analytical study and comparison.


\sscn{Elliptic model}{s:casee}

Let us consider a special case when $\beta = \alpha$ (for the original non-Hermitian
parameters it reads $\beta = \alpha m^2
\omega^2$),
then Eq. (\ref{mbhamg-qp}) takes the form
\ba
\hat H
&=&
\frac{\hat p^2}{2}
+
\frac{\hat q^2}{2}
,\ \
\hat\Gamma
=
\frac{\alpha}{2}
\left(
\hat p^2
+
\hat q^2
\right)
+
\frac{\gamma}{2}
,
\lb{mbhamg-qpo2}
\ea
or,
alternatively,
\ba
\hat H
=
\hat{a}^\dagger \hat{a}
+ \frac{1}{2}
,
\ \
\hat\Gamma
=
\alpha
\left(
\hat{a}^\dagger \hat{a}
+ \frac{1}{2}
\right)
+
\frac{\gamma}{2}
,
\ea
while the total Hamiltonian is given by Eq. (\ref{e:nhhtot}), as
before. 
Such models often appear in quantum optics, where the real component of the 
coefficient in front of the number operator 
$\hat N = \hat{a}^\dagger \hat{a}$ describes Rabi-type oscillations,
whereas the non-Hermitian component 
effectively describes concomitant dissipative effects, see Ref. \cite{sz14}, and Ch. 6 of \cite{bpbook}.

Note that in this case the total Hamiltonian is a normal
operator, therefore operators
$\hat H$ and $\hat\Gamma$ commute and thus have a common
set of eigenfunctions.

There is a natural O(2) action on
 $\mathbb{R}^2$, moreover, for $\alpha=\beta$, the right-hand-side of
 Eq. (\ref{e:evgen}) is invariant under this action. Consequently,
polar coordinates, defined by orbits of action of the group O(2),
are convenient for analysis of Eq. (\ref{e:evgen}), which can be
rewritten  as
\ba
\frac{\p W}{\p t}=
\frac{\alpha}{4}
\Delta^{\!(2)} W-(\alpha R^2 +\gamma) W
+ \frac{\partial}{\partial \Phi}  W
,
\label{e41}
\ea
where
$W=W (R,\Phi,t)$,
$
\Delta^{\!(2)}
=
\frac{\p^2 }{\p p^2} + \frac{\p^2 }{\p q^2}
=
\frac{\p^2 }{\p R^2}
+\frac{1}{R}
\frac{\p }{\p R} +\frac{1}{R^2}\frac{\p^2 }{\p \Phi^2}$ is a 2D phase-space
Laplacian,  $R = \sqrt{q^2 + p^2}$ is a  (dimensionless)
phase-space radius; $\Phi$ is a phase-space angle defined as $\Phi
={\arctan} (p/q)$, for {\mbox{$p>0$}}, {\mbox{$q>0$}}, and by its
standard modifications in other quadrants.
Equation~(\ref{e41})
must be supplemented with the following conditions
\ba
&& W (R,\Phi,t) \in \mathbb{R} ,\lb{e:fdsbc0}
\\&&
\lim\limits_{R\to 0_+} |W (R,\Phi,t)| < \infty ,
\lb{e:fdsbc1}
\\
&& \exists \delta > 2:\!
 \lim\limits_{R\to +\infty} R^\delta W (R,\Phi,t) = 0 ,
\lb{e:fdsbc2}
\\&&
W (R,\Phi + 2 \pi,t) = W (R,\Phi,t)
,
\lb{e:fdsbc3}
\ea
which are required by the
probabilistic meaning of $W$.
In order to
implement an assumption that the dissipation switches on at
$t=0$,
but until then the density is normalized to one,
these conditions must be considered together
with the initial-time normalization:
\be\lb{e:fdsbc4}
\int
W (R,\Phi,0) R \, d R \, d\Phi
=
1
,
\ee
where integration is taken over the whole phase space.

A wide range of solutions of Eq. (\ref{e41}) satisfying conditions
(\ref{e:fdsbc0})-(\ref{e:fdsbc4}) can be constructed as linear
combinations
of the basis functions
\ba
&&B_{\lambda,\nii} (R,\Phi,t)=  e^{-\lambda t} e^{i \nii \Phi} b
(R),
\label{ansatz}
\ea
where
$\lambda$ and $\nii$ are 
constants whose meaning and range will be specified shortly.
The
radial part of Eq. (\ref{ansatz})
must obey conditions
\ba
&&\lim\limits_{R\to 0_+} |b(R)|  < \infty ,  \lb{e:wbc2}
\\
&&\exists \delta > 2: \!\lim\limits_{R\to +\infty} R^{\delta}
b(R)  = 0
,
 \lb{e:wbc3}
\ea
as required by Eqs. (\ref{e:fdsbc1}) and (\ref{e:fdsbc2}).

Boundary conditions (\ref{e:fdsbc0})-(\ref{e:fdsbc3}) pose an
eigenvalue problem, which limits a range of allowed values of
$\lambda$
 and $\nii$.
For instance,
the condition (\ref{e:fdsbc3}) immediately imposes that values of
$\nii$ in Eq. (\ref{ansatz}) are limited to integers.
Substituting Eq.~(\ref{ansatz}) into (\ref{e41}),  we obtain 
\be
b''(R)+
\frac{1}{R} b'(R)
-\Big[ 4R^2-\frac{4(\lambda-\gamma+i\nii)}{\alpha}  +
\frac{\nii^2}{R^2}\Big]
b(R)
=0 ,
\label{e26}
\ee
where prime means a derivative with respect to $R$.

In general, the constant $\lambda$ can be complex-valued, therefore
we can write it in a form $\lambda=\lambda_r + i\lambda_i$, for
$\lambda_r,\lambda_i\in
\mathbb{R}$.
Nontrivial solutions of Eq. (\ref{e26}) fulfill boundary conditions
(\ref{e:wbc2}) and (\ref{e:wbc3}) iff $
\lambda_i = \Im (\lambda) = -\nii
$ and 
$
\frac{1}{2\alpha} (\lambda_r -\gamma)-\frac{1}{2} (1+|\nii|)
$ is a natural
number,
which
imposes quantization conditions on $\lambda_r $.
Solutions of Eq. (\ref{e26}),
satisfying these conditions, can be expressed as
\ba
b_{n,\nii}(R)=
\frac{(-1)^n}{\pi} R^{|\nii|} e^{-R^2} L_n^{|\nii|}(2R^2),
\ \
\label{e18}
\ea
where
$L_n^{|\nii|}(z)$ is the generalized Laguerre polynomial.
For a given $\nii$, real-valued functions
(\ref{e18}) form an orthogonal set in a space of all square integrable
functions on a positive half-axis with measure $RdR$.
From Eqs. (\ref{ansatz}) and (\ref{e18}), we obtain
\be
B_{n,\nii} (R,\Phi,t)
=
\frac{(-1)^n}{\pi}
e^{-\lambda t} e^{i \nii \Phi} R^{|\nii|} e^{-R^2}
L_n^{|\nii|}(2R^2)
,\label{e20}
\ee
where 
$\lambda$ takes values from a discrete complex
spectrum,
\be
\lambda =
\lambda_{n,\nii} = \alpha(2n+1+|\nii|)+\gamma-i \nii
.\ee
Correspondingly,
the
real part of $\lambda$,
\be\label{spectr}
\tau_{n,\nii}^{-1}
=
\Re(\lambda_{n,\nii})
=
\alpha(2n+1+|\nii|)+\gamma,
\ee
yields spectral values of the decay constant (inverse mean lifetime)
of the system (\ref{mbhamg}), while the imaginary one contributes to
the phase.
For $\nii=0$, functions (\ref{e20}) correspond to projection on 
$n$th eigenstate of Hamiltonian. For $\nii\neq 0$, they 
correspond to transition between states for which energy levels 
differ by $\nii$.

From a physical viewpoint, appearance of a discrete spectrum can
be easily explained by writing Eq. (\ref{e26}) in a Schr\"odinger
form. Using a transformation
\be
b (R) = \psi (R)/\sqrt R,
\ee
one can rewrite
Eq. (\ref{e26}) as a stationary Schr\"odinger equation for a
fictitious particle of mass $1/2$ and energy $4(\lambda_r -
\gamma)/\alpha$, moving in the external potential $U (R) = 4 R^2 +
(\nii^2-1/4)/R^2$. The shape of the latter, along with the conditions
(\ref{e:wbc2}) and (\ref{e:wbc3}), indicates an existence of bound
states.


A quantum number $n$ is the same as the one in the initial energy spectrum 
\be\lb{spectrE}
E^{(0)}_n
=
n + 1/2
,
\ee
which occurred before the interaction with environment was switched on,
due to the quantization of the Hermitian part of Eq. (\ref{e:nhhtot});
this identification is a result of the normality condition discussed after Eq. (\ref{mbhamg-qpo2}).
Therefore, the quantization of mean lifetime
is additional to this energy eigenvalue quantization.
By analogy with an interpretation of conventional quantum spectra,
one can say that quantum dissipation can,
in some sense, restrict itself -- by creating forbidden ranges of
decay constants'  values and forming 
the ``dissipative bound'' states. 
Small disturbances caused by interaction with an environment will
not cause a system to make a transition and leave such states,
therefore it is natural to expect that long-lived
states will exist even in a slightly perturbed system,
as will be demonstrated below.

\begin{figure}
\centering
\subfloat[$B_{0,0}^+$]{
  \includegraphics[width=\sct\columnwidth]{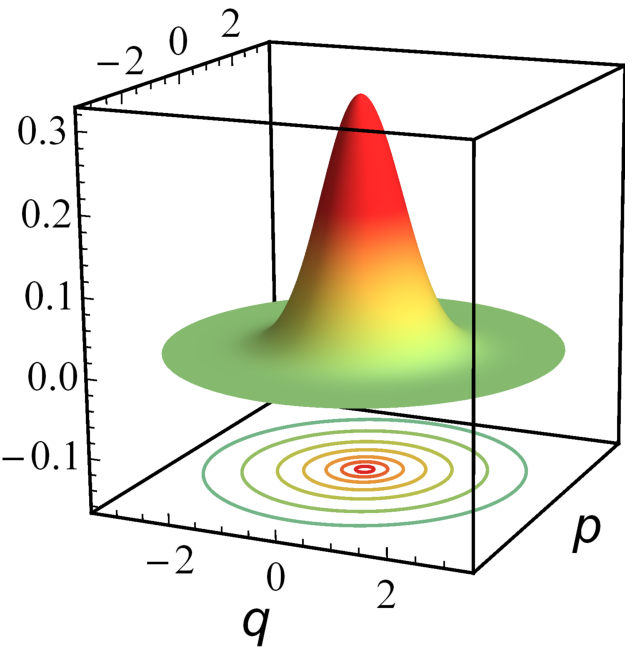}
}
\subfloat[$B_{0,1}^+$]{
  \includegraphics[width=\sct\columnwidth]{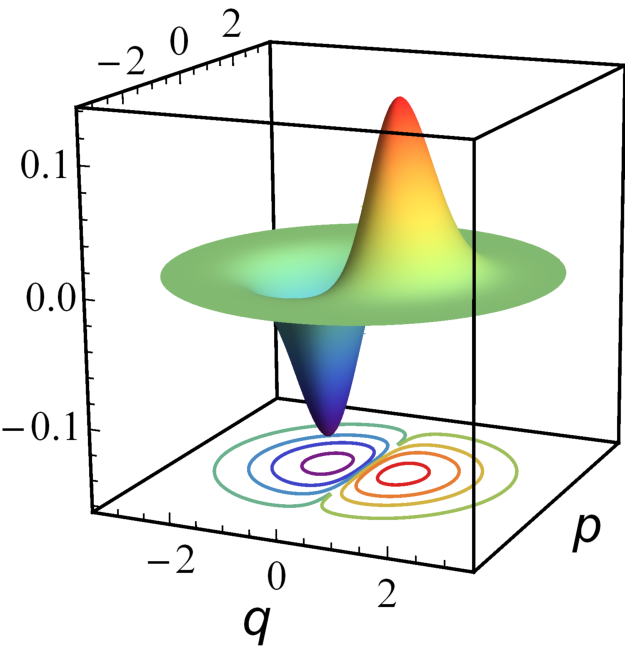}
}
\hspace{0mm}
\subfloat[$B_{0,2}^+$]{
  \includegraphics[width=\sct\columnwidth]{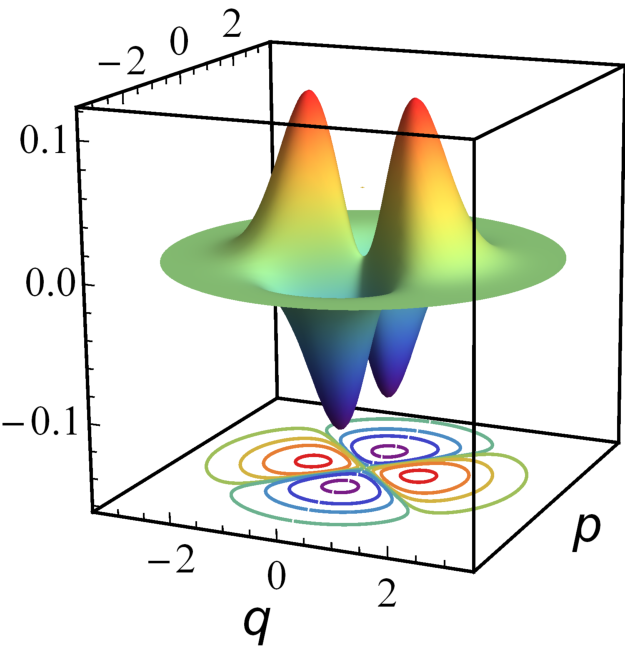}
}
\subfloat[$B_{0,3}^+$]{
  \includegraphics[width=\sct\columnwidth]{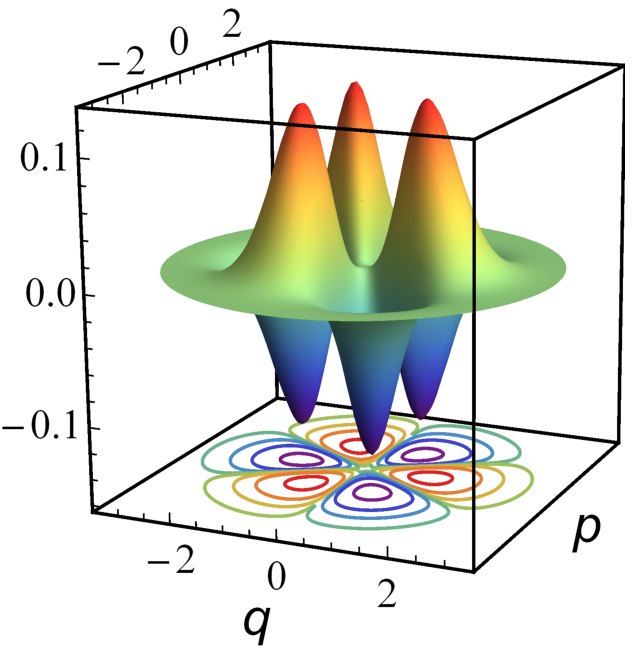}
}
\caption{Phase-space plots of functions (\ref{sol1}) for $n=0$ states,
evaluated at $t=0$
and
$\alpha = 1$, $\gamma = 0$.
}
\label{f:w0}
\end{figure}

Furthermore, in order to satisfy the reality condition
(\ref{e:fdsbc0}), we have to construct real-valued linear
combinations of basis functions (\ref{e20}). For example, it is easy to
check that the following functions satisfy the eigenproblem
(\ref{e41})-(\ref{e:fdsbc3}):
\ba
B_{n,\nii}^+ (R,\Phi,t) &\equiv&
\frac{1}{2}
\left[B_{n,\nii} (R,\Phi,t)+B_{n,-\nii} (R,\Phi,t)
\right]
\nn\\&=&
A (t,R)
\cos{[\nii(t+\Phi)]},
\label{sol1}\\
B_{n,\nii}^- (R,\Phi,t) &\equiv&
\frac{1}{2 i}
\left[B_{n,\nii} (R,\Phi,t) - B_{n,-\nii} (R,\Phi,t)
\right]
\nn\\&=&
A (t,R)
\sin{[\nii(t+\Phi)]},
\label{sol2}
\ea
where the amplitude is given by
\be
A (t,R) = \frac{(-1)^n}{\pi} e^{-t/\tau_{n,\nii}} R^{|\nii|}
e^{-R^2}
L_n^{|\nii|}(2R^2)
.
\ee
Note that from now on it suffices to consider
only natural values of $\nii$.
Functions $B_{n,\nii}^+ (R,\Phi,0)$ and
$B_{n,\nii}^- (R,\Phi,0)$ form a basis, in which an
initial Wigner function can be expanded,
for one could
determine its time evolution.

\begin{figure}
\centering
\subfloat[$B_{1,0}^+$]{
  \includegraphics[width=\sct\columnwidth]{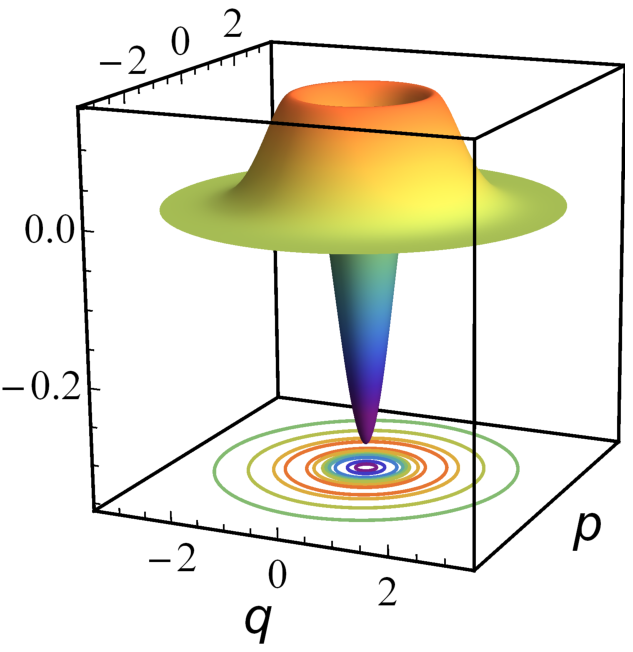}
}
\subfloat[$B_{1,1}^+$]{
  \includegraphics[width=\sct\columnwidth]{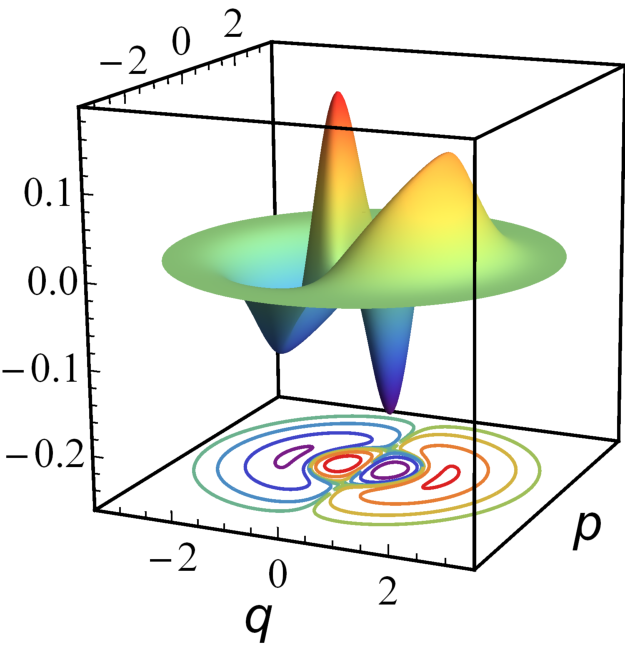}
}
\hspace{0mm}
\subfloat[$B_{1,2}^+$]{
  \includegraphics[width=\sct\columnwidth]{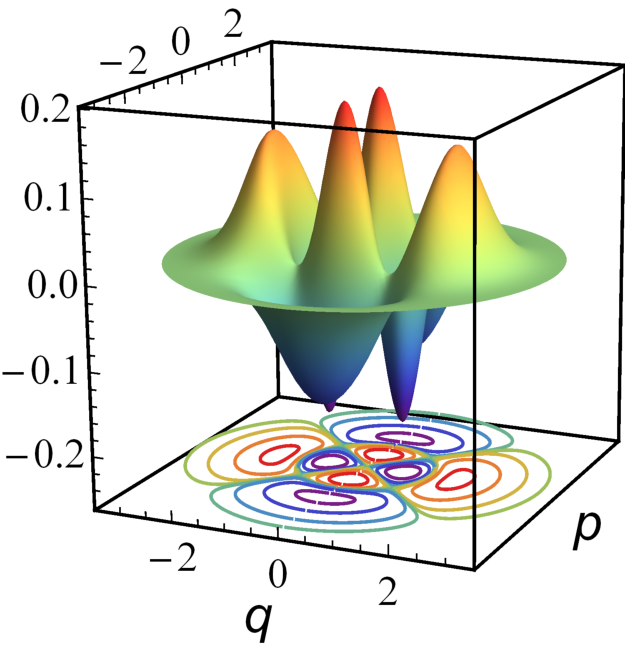}
}
\subfloat[$B_{1,3}^+$]{
  \includegraphics[width=\sct\columnwidth]{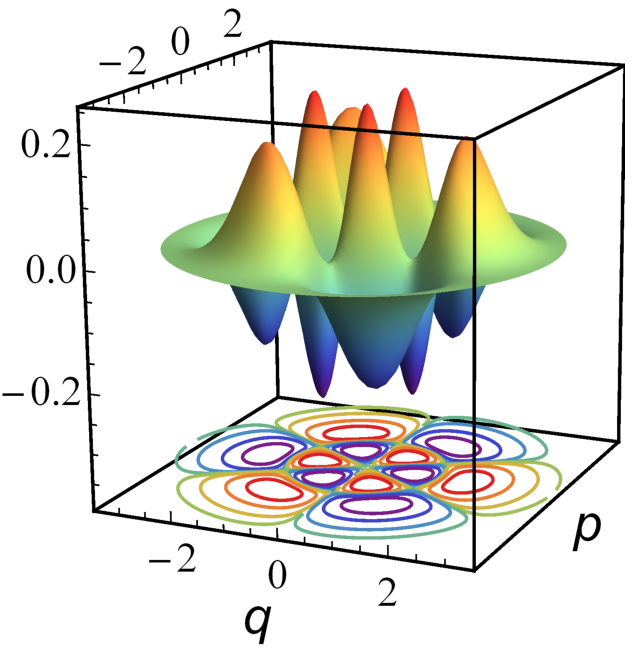}
}
\caption{Phase-space plots of functions (\ref{sol1}) for $n=1$ states,
evaluated at $t=0$
and
$\alpha = 1$, $\gamma = 0$.
}
\label{f:w1}
\end{figure}
\begin{figure}
\centering
\subfloat[$B_{2,0}^+$]{
  \includegraphics[width=\sct\columnwidth]{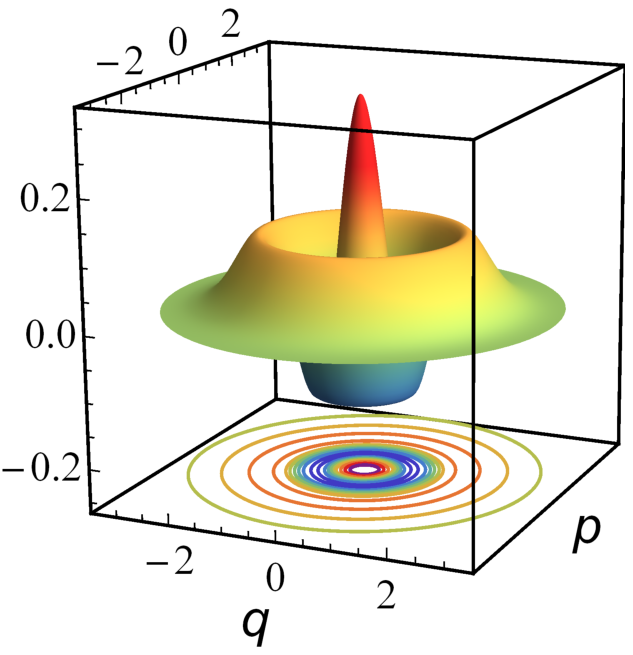}
}
\subfloat[$B_{2,1}^+$]{
  \includegraphics[width=\sct\columnwidth]{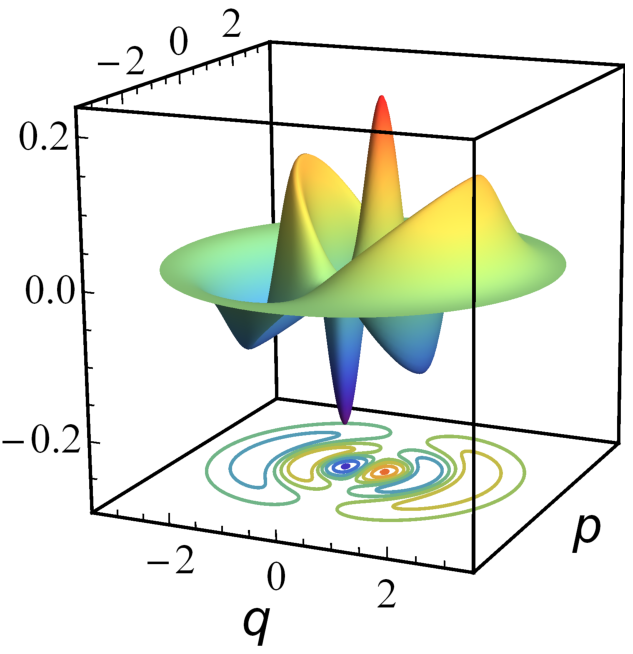}
}
\hspace{0mm}
\subfloat[$B_{2,2}^+$]{
  \includegraphics[width=\sct\columnwidth]{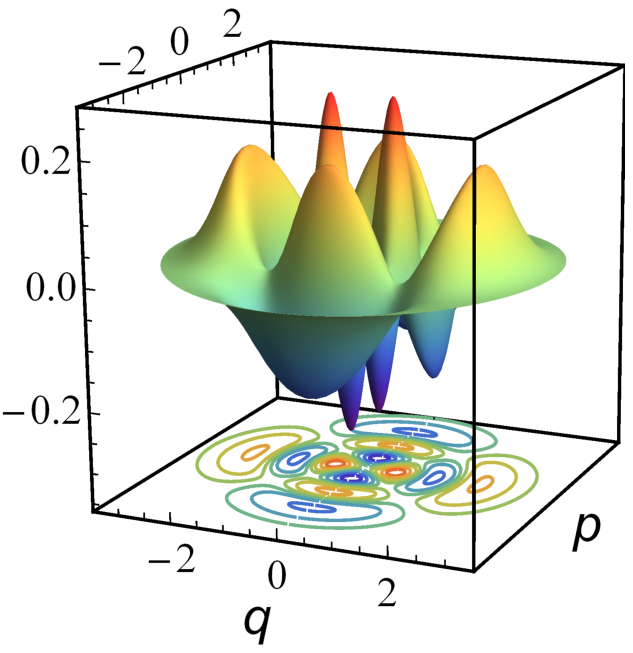}
}
\subfloat[$B_{3,0}^+$]{
  \includegraphics[width=\sct\columnwidth]{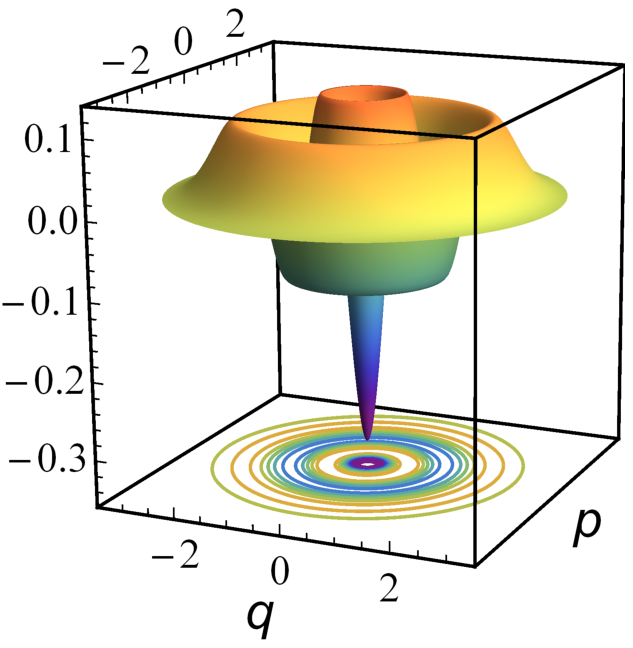}
}
\hspace{0mm}
\subfloat[$B_{3,1}^+$]{
  \includegraphics[width=\sct\columnwidth]{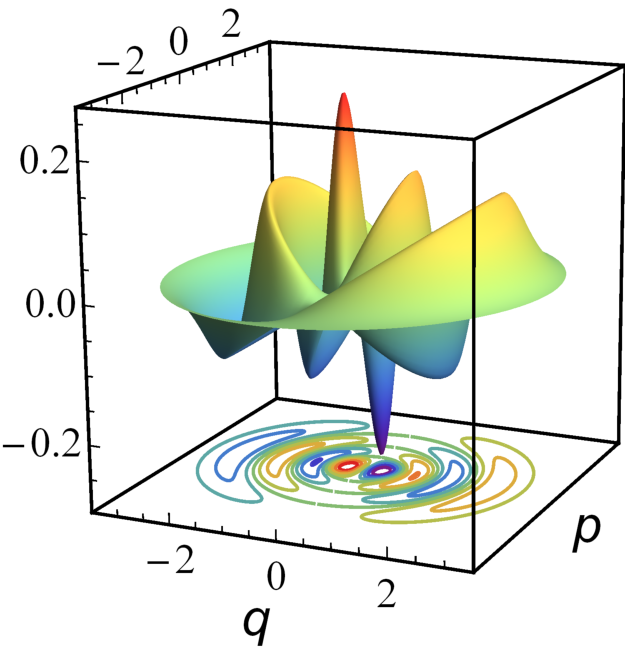}
}
\subfloat[$B_{3,2}^+$]{
  \includegraphics[width=\sct\columnwidth]{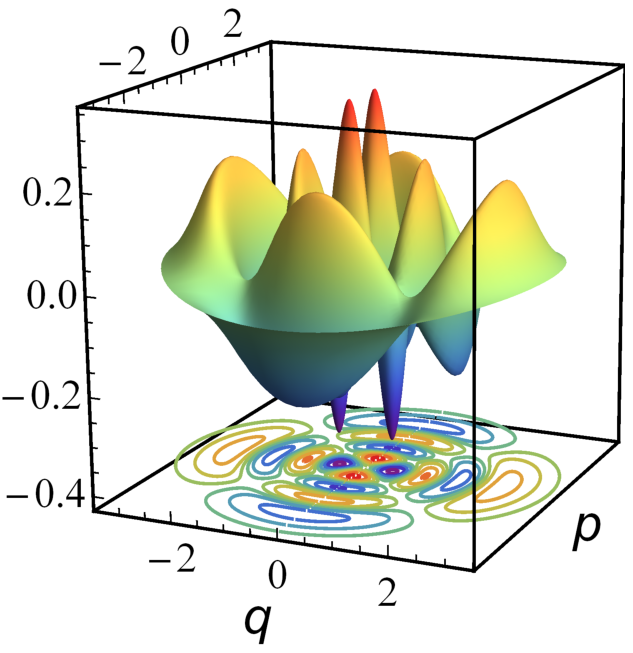}
}
\caption{Phase-space plots of functions (\ref{sol1}) for $n=2,3$ states,
evaluated at $t=0$
and
$\alpha = 1$, $\gamma = 0$.
}
\label{f:w23}
\end{figure}

In order to visualize functions (\ref{sol1}) and
(\ref{sol2}), notice that the time evolution not only decreases
their initial amplitudes exponentially (except long-lived states discussed below)
but also rotates them in a
$qp$-plane.
Moreover, one can check that $B_{n,\nii}^+$ and
$B_{n,\nii}^-$ can be transformed into each other by an appropriate
phase shift of the polar angle (except $B_{0,0}^-$ which is
identically zero).
Therefore, for illustration purposes, it suffices to
consider only the functions $B_{n,\nii}^+$ at the initial moment of time,
so plots
of some of them are given in Figs.
\ref{f:w0}--\ref{f:w23}. One can see that, for $|t|<\infty$, all the functions $B_{n,0}^+$
are surfaces of revolution obtained by revolving a curve with $n+1$
local extrema.
For larger $\nii$'s, rotational symmetry breaks down and
 functions $B_{n,\nii}^+$ have $2\nii(n+1)$ local extrema.

Note that not all basis functions alone can be
used for representing a density operator in this approach
because not all of them fulfill the condition (\ref{e:fdsbc4}).
One can deduce that expansion of a
Wigner function into series of $B_{n,\nii}^+ $ and $B_{n,\nii}^-$ must always
contain at least one term with $\nii=0$, because only these terms
have a non-zero contribution to normalization (terms with $\nii\neq
0$ that do not contribute to normalization are, however, essential
for calculating average values of observables that depend on $\Phi$).



Combining Eqs. (\ref{spectr}) and (\ref{spectrE}), we obtain the 
constraint
\be\label{spectr00}
\alpha
\left(
2 E^{(0)}_n + |\nii| +
\gamma/\alpha
\right)
\tau
=
1
,
\ee
which indicates that energies and mean lifetimes cannot be independent
here.
From this, one can immediately derive the resonance condition
\be
E^{(0)} \to E^{(0)}_c = - \frac{1}{2}
\left(
|\nii|
+
\frac{\gamma}{\alpha}
\right)
,
\ee
at which the mean lifetime reaches a maximum value as a function of the initial energy.
If the non-Hermitian parameters happen to obey also
the condition
\be
\left(
\gamma/\alpha
\right)_c
=
- 2 k - 1
- |\nii|,
\ee
$k$ being any natural number,
then there exists a state, with the quantum number
\be
n_c =
- \frac{1}{2}
(1+ |\nii|
+
\gamma/\alpha)
,
\ee
which
acquires an infinite mean lifetime.
This state is stationary
but only in the absence of quantum fluctuations or additional interactions, otherwise
it is a long-lived one, with a finite width.
This can be used to explain the existence of long-lived states in systems with
quasi-continuous energy spectra, such as atoms in the photon or phonon bath in potential wells.
In the former case, the long-lived state of an electron acquires a finite width due to its interaction with photons, and in the case where an electron is ``dressed'' by phonons, the long-lived polaronic state forms.

From the point of view of the formalism itself, the model provides 
an example of a non-Hermitian Hamiltonian system whose states can be 
both decaying (or unstable) and long-lived,
depending on a state's quantum number.
The existence of NH-driven long-lived states suggests that quantum 
dissipative effects do not always lead to decay or critical 
instability of a system at large times, but can sustain its 
stability, under certain conditions. Notice that for the 
above-mentioned state $n = n_c$, the time exponent disappears, which 
is equivalent to considering the sustainable type of evolution (the 
latter would be governed by the normalized Wigner density function 
which equals to $A (0, R)$ in this case). This confirms that 
continuous normalization (\ref{e:rho}) is instrumental in describing 
those physical processes for which one needs to automatically 
dispose decaying or unstable states,
but leave the long-lived ones.

In eigenfunctions (\ref{e20}), one
can separate a time part from the position variables: 
$B_{n,\nii}(R,\Phi,t)=P_{n,\nii}(R,\Phi)T_{n,\nii}(t)$,
where
$T_{n,\nii}(t)=\exp[
-(
\tau_{n,\nii}^{-1} 
-i\nii) t
]$. 
In order to find an energy spectral 
decomposition of $T_{n,\nii}(t)$,
we consider its 
Fourier transform.
We take into account here that for 
$
\tau_{n,\nii} 
\neq 0$ 
the Fourier 
integral does not exist; 
for 
$
\tau_{n,\nii} 
= 0$ it exists 
only in distributional sense. 
Thus, depending on a sign of 
$\tau_{n,\nii}$, we will consider integrals over positive or 
negative half-axes. 
For $
\tau_{n,\nii} 
>0$, we obtain the transform
\ba
F^{+}_{n,\nii}(E)
=\int_0^\infty e^{iEt} T_{n,\nii}(t)dt
=
\left[i(E+l)+
\tau_{n,\nii}^{-1}
\right]^{-1}\!.~~
\label{Fplus}
\ea
On the other hand,
the energy space probability distribution function 
can be computed
as 
$
f (E)
=
{\cal N}_f
\left|
\int_{\cal T} W \, e^{iEt}  dt
\right|^2
$, 
where
${\cal T}$ is a suitably chosen domain of integration,
and
${\cal N}_f$ is the factor determined by the normalization condition
$\int_{-\infty}^{\infty} f(E) \, d E = 1$.
Therefore,  
in our case
this function acquires
the 
Cauchy-Lorentz-Breit-Wigner form:
\be
f_{n,\nii}(E)
=
\frac{1}{\pi} \Upsilon_{n,\nii}
|F^+_{n,\nii}(E)|^2
=
\frac{1}{\pi}
\frac{\Upsilon_{n,\nii}}{
(E+l)^2+
\Upsilon_{n,\nii}^2
}
,
\label{fplus}
\ee
where 
$\Upsilon_{n,\nii} = \tau_{n,\nii}^{-1}$.
This formula 
describes a resonance 
with HWHM scale parameter 
$\Upsilon_{n,\nii}$ 
and location parameter $\nii$ that corresponds 
to energy difference between eigenfunctions of Hamiltonian 
(\ref{mbhamg-qpo2}) separated by $\nii$  energy levels. 
When the parameter $\gamma$ varies in such a way that 
$\Upsilon_{n,\nii}\rightarrow 0$, 
then \eqref{fplus} converges to $\delta(E+l)$, in a
distribution sense.

\sscn{
Hyperbolic model}{s:solh}

Let us consider a case when $\beta = - \alpha$ (for the original non-Hermitian
parameters it reads $\beta = - \alpha m^2
\omega^2$),
then Eq.~(\ref{mbhamg-qp}) takes the form
\ba
\hat H
&=&
\frac{\hat p^2}{2}
+
\frac{\hat q^2}{2}
,\ \
\hat\Gamma
=
\frac{\alpha}{2}
\left(
\hat p^2
-
\hat q^2
\right)
+
\frac{\gamma}{2}
,
\lb{mbhamg-qpo3}
\ea
or,
alternatively,
\ba
\hat H
=
\hat{a}^\dagger \hat{a}
+ \frac{1}{2}
,
\ \
\hat\Gamma
=
- \frac{\alpha}{2} \left( \hat{a}^{\dagger 2} + \hat{a}^2 \right)
+
\frac{\gamma}{2}
,
\ea
whereas Eq. (\ref{e:evgen}) simplifies to
\ba
\frac{\partial}{\partial t} W
&=&
-\alpha
\left(
 p^2
-
 q^2
+ \frac{\gamma}{\alpha}
\right)
W
-
\left(
p \frac{\partial}{\partial q}
-
q \frac{\partial}{\partial p}
\right)
W
\nn\\&&
+
\frac{\alpha}{4}
\left(
 \frac{\partial^2}{\partial q^2}
-
\frac{\partial^2}{\partial p^2}
\right)
W
,
\lb{e:evgenhb}
\ea
where
$
W = W (q, p, t) =
e^{- \lambda t}
W(q, p)
$. 
This is a partial differential equation of a hyperbolic type,
whose solutions are substantially more complicated, hence their discussion
is beyond a scope of this paper.
Here we just note that the square integrable
solutions are not orthogonal to each other although one can form
a (non-orthogonal) basis in $L^2$ from them.

As for the spectrum then it can still be  obtained in a simple form:
\be
\lambda=\lambda_\nii
=
\gamma +  i \nii \sqrt{1+\alpha^2}
,
\ee
where $\nii\in \mathbb{Z}$. Therefore, for the hyperbolic-type 
non-Hermitian harmonic oscillator, the only physically allowed mean 
lifetime's eigenvalue would be simply
\be
\tau^{-1}
=
\gamma
,
\ee
which thus depends neither on any quantum number
nor on $\alpha$, unlike its elliptic counterpart \eqref{spectr}.

Finally, 
considerations related to the spectral 
decomposition of a time-dependent part of solutions of Eq.
\eqref{e:evgenhb},
written in the form $e^{-\lambda t}W(q,p)$, lead to the energy
distribution 
of a type \eqref{fplus}:
\be
f (E)
=
\frac{1}{\pi}
\frac{\Upsilon}{
(E - \bar E)^2+
\Upsilon^2
}
,
\label{e:pdfh}
\ee
with the HWHM and location parameters being
$
\Upsilon 
=
\tau^{-1}
=\gamma
$
and
$
\bar E = \nii\sqrt{1+\alpha^2}
$,
respectively. 
Contrary to the elliptic case, 
$\Upsilon$ does not depend on a state here,
and 
for a given value of 
$\gamma$, the variation of $\nii$ does not change shape of this distribution 
but corresponds to a translation parallel to the $E$ axis. 


To conclude, the considered cases illustrate that values of non-Hermitian parameters
strongly affect time evolution of a system.

\scn{Conclusion}{s-con}

We apply the Wigner-Weyl transform and phase space formulation for the evolution equation of density operator in systems described by non-Hermitian Hamiltonians, such as the non-Hermitian quantum harmonic oscillator. It turns out that the Wigner quasiprobability distribution functions for such systems become eigenfunctions of an emergent eigenproblem of the Sturm-Liouville type, which leads to a discrete spectrum's appearance.

It is thus shown that in presence of the dissipative environment of a certain type, mean lifetime and decay constants do not necessarily take arbitrary values but obey a discrete eigenvalue spectrum. This effect is somewhat analogous to the quantization of energy spectrum of bound states in conservative quantum systems.

In order to illustrate our approach,
in Sec. \ref{s:cases}
we have considered two exactly-solvable models of non-Hermitian harmonic oscillators,
which
exhibit a different behavior depending on values of the non-Hermitian parameters.
We applied these models for describing those unstable particles and resonances
which can be modeled by a quadratic Hamiltonian with respect to positions and momenta.
The corresponding Breit-Wigner distribution parameters have been derived and studied.

One may ask
whether the discussed
effects would still exist in open quantum systems that are more
general than those described by the non-Hermitian harmonic
oscillator.
For
those anharmonic systems, the evolution equation (\ref{er}) would still be
a linear differential equation of an order equal or larger than two.
Since discrete-spectrum eigenvalue problems abundantly appear in
differential equations of a second order and above, under boundary
conditions of a type (\ref{e:fdsbc1})-(\ref{e:fdsbc3}), the proposed
mechanism must hold for a large class of non-Hermitian models.

\begin{acknowledgments}
The research of K.Z. is supported by the National Research
Foundation of South Africa under Grants Nos. 95965 and 98892.
Proofreading of the manuscript by P. Stannard is greatly appreciated.
\end{acknowledgments}


\end{document}